\documentclass[twocolumn,aps,prl,showpacs]{revtex4}
\usepackage{graphicx}
\usepackage{amssymb}
\usepackage{bm}
\usepackage{amsmath}

\makeatletter

\begin{document}

\title{Emergence of Plasmaronic Structure in the Near Field Optical Response of Graphene}
\author{J. P. Carbotte$^{1,2}$}
\author{J. P. F. LeBlanc$^{3,4}$}
\email{jpfleblanc@gmail.com}
\author{E. J. Nicol$^{3,4}$}

\affiliation{$^1$Department of Physics and Astronomy, McMaster
University, Hamilton, Ontario L8S 4M1 Canada}
\affiliation{$^2$The Canadian Institute for Advanced Research, Toronto, ON M5G 1Z8 Canada}
\affiliation{$^3$Department of Physics, University of Guelph,
Guelph, Ontario N1G 2W1 Canada} 
\affiliation{$^4$Guelph-Waterloo Physics Institute, University of Guelph, Guelph, Ontario N1G 2W1 Canada}

\date{\today}
\begin{abstract}
The finite momentum optical response $\sigma({\boldsymbol{q}},\omega)$ of graphene can be probed with the innovative technique of infrared nanoscopy
where mid-infrared radiation is confined by an atomic force microscope cantilever tip.  
In contrast to conventional $q\sim 0$ optical absorption which primarily involves Dirac fermions with momentum near 
the Fermi momentum, $k\sim k_F$, for finite $q$, $\sigma({\boldsymbol{q}},\omega)$ has the potential to provide information on many body renormalizations and collective phenomena which have been found
at small $k< k_F$ near the Dirac point in electron-doped graphene. For electron-electron interactions, the low energy excitation spectrum characterizing the incoherent part of the quasiparticle spectral function
 of  Dirac electrons with $k\sim k_F$ consists of a flat, small amplitude background which  scales with chemical potential and Fermi momentum.
However, probing of the states with $k$ near $k=0$ will reveal
plasmarons, 
a collective state of a charge carrier and a plasmon.  These collective modes in graphene
 have recently been seen in angle-resolved photoemission spectroscopy 
and here we describe how they manifest in near field optics.
\end{abstract}
\pacs{78.67.Wj,73.22.Pr,73.20.Mf}

\maketitle

\section{Introduction}
Since its discovery, graphene continues to provide a rich trove of exotic 
phenomena arising out of novel charge dynamics which map onto a Hamiltonian
for massless Dirac fermions, providing a low energy bandstructure of two
linear bands which touch and cross at a so-called Dirac point.
Using angular resolved photoemission spectroscopy (ARPES) Bostwick at al. \cite{bostwick:2010} have recently observed, 
in samples of graphene grown epitaxially on a H-SiC substrate, that the Dirac point splits into two at momentum $k=0$.  
A plasmaron ring is observed between the two Dirac points which has its origin in electron-electron interactions 
(EEI)\cite{hwang:2008,polini:2008,sensarma:2011} and consequently is not part of the bare band description.
In metals, the long range nature of the Coulomb interactions provides the mechanism for the formation of long-lived collective 
charge density oscillations or plasmons.  These also exist in graphene\cite{hwang:2007, sensarma:2010} although with an altered 
dispersion relation, {\it i.e.,} in the long wavelength limit ($q \to 0$) there is a square root dependence on $q$.  
As recognized by Lundquist\cite{lundqvist:1967}, charged quasiparticles can interact with the plasmons to form a composite state, termed a plasmaron.  
It is these composite excitations that provide an understanding of the ARPES data in the region around $k\simeq 0$ near the split Dirac point.  
Other evidence of plasmarons in graphene might be found in the tunneling conductance. The observed\cite{zhang:NP:2008, brar:2010} lifting of the node in the bare band density of
quasiparticle states provides evidence of the presence of many-body effects\cite{nicol:2009, carbotte:2010, leblanc:2011, principi:2011}, moreover plasmaronic structures in this region,
which scale with chemical potential,\cite{leblanc:2011, principi:2011}, 
may underlie the EEI-type features identified\cite{brar:2010} in tunneling experiments.
Aside from graphene, the elusive plasmaron has been observed in other systems \cite{shay:1971} such as elemental Bi semimetals.\cite{tediosi:2007, armitage:2010} 
   
\begin{figure}
  \begin{center}
  \includegraphics[width=0.8\linewidth]{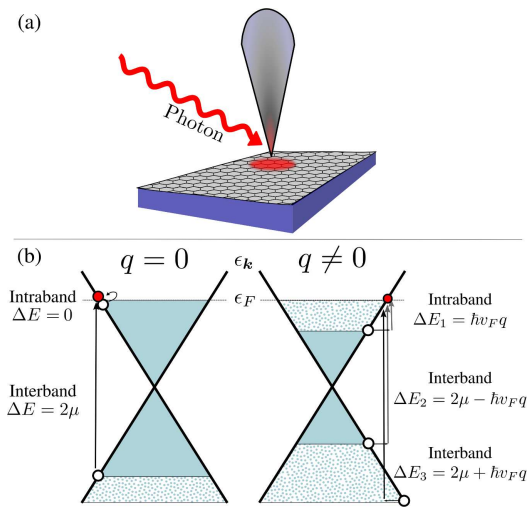}
  \end{center}
  \caption{\label{fig:1}(Color online) (a) Schematic of a graphene sheet on a substrate with an AFM tip used to confine the incident radiation to a nanoscale region around 
the tip as in  Fei et al.\cite{fei:2011}.  (b)  Schematic of intraband and interband transitions in the Dirac cone bandstructure of 
graphene. Transitions from occupied (shaded) to unoccupied states for $q=0$ photon absorption are shown on the left with the heavy shading indicating the Pauli-blocked region.
For a finite $q$ transfer, new optical absorption processes (shown at right) are possible. For example, a particle at $k=k_F-q$ may be  excited  
to $k=k_F$ with finite $q$ transfer. The Pauli-blocked region is reduced as $q$ increases and the plasmaron region around $k=0$ can be unveiled and probed
by optics.}
\end{figure}

Optical spectroscopy provides useful information on charge carrier dynamics and has widely been used to probe the nature of the Dirac quasiparticles in graphene.  
Many experimental results are reviewed in the paper of Orlita and Potemski.\cite{orlita:2010}  While most of the observations can be understood qualitatively and even quantitatively 
with bare bands, some cannot.  A prominent example is the absorption spectrum of graphene as described by the real part of its AC conductivity at $q=0$, $\sigma(\boldsymbol{q}=0,\omega)$.  
Theory predicts a Drude peak centered about zero photon energy ($\omega$) followed by a region of near zero absorption
before a rapid rise to the universal value of $\sigma_0=\pi e^2/2h$ at $\omega=2\mu$, where $\mu$ is the chemical potential.\cite{gusynin:2007,gusynin:2009, peres:2008} 
This background remains nearly constant over a large energy range above $2\mu$.  The Drude piece comes from intraband absorption [see lefthand side of Fig.~\ref{fig:1}(b)]
and its width is set by the residual scattering rate $\gamma$ assumed to be much smaller than $2\mu$.  The interband optical transitions, which provide the universal background, begin only 
at $\omega=2\mu$ because of Pauli blocking, {\it i.e.}, Pauli exclusion blocks vertical transitions for photon absorption below this energy due to an occupied
final state for $k<k_F$ [Fig.~\ref{fig:1}(b), $q=0$].  
One aspect of this bare band picture is not confirmed in experiment.  The conductivity never falls below about $\sigma_0/3$ in the Pauli-blocked region of the spectrum 
where it would be expected to be almost zero.\cite{li:2008}  This has been widely interpreted as evidence for the importance of Coulomb correlations and 
the electron-phonon interaction which provides phonon-assisted Holstein sidebands\cite{carbotte:2010, stauber:2008b} to the main Drude peak.  
A second contribution comes from equivalent sidebands associated with the rise of the interband optical transitions at $\omega=2\mu$.\cite{carbotte:2010}
 
In this letter, we address the issue of how plasmaronic structures present themselves in the optical conductivity of graphene.  Conventional experiments measure  
the $q\sim 0$, long wavelength limit, of the AC conductivity $\sigma(\boldsymbol{q},\omega)$.  For this case the intraband transitions mainly involve 
the quasiparticle states at and near momentum $k=k_F$ in the top Dirac cone, while the interband transitions involve $k\geq k_F$ for 
both upper and lower Dirac cones [lefthand side of Fig.~\ref{fig:1}(b)].
Consequently, the plasmaron region found around $k\simeq 0$ (as we have described) is not directly probed and one does not expect to see a significant 
signature of the plasmarons in the $q\sim 0$ AC conductivity.
However, this is not the case for the finite $q$ conductivity. 

\begin{figure}
  \begin{center}
  \includegraphics[width=0.8\linewidth]{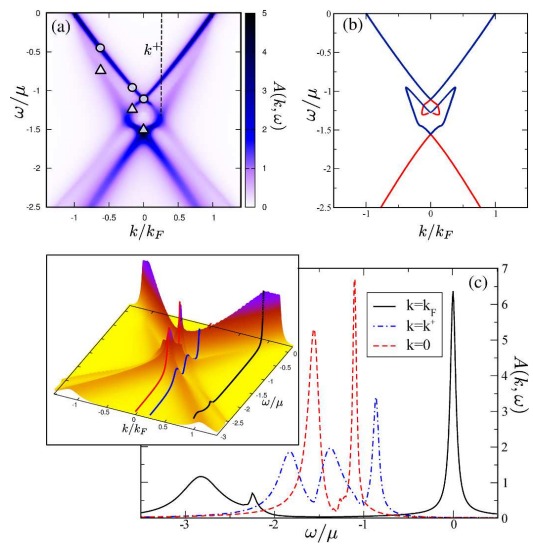}
  \end{center}
  \caption{\label{fig:2}(Color online) (a) Color map of the Dirac fermion spectral function $A(k,\omega)$ including EEI. 
   (b) Solutions of Eq.~(\ref{eqn:poles}) for  quasiparticle energies dressed by EEI in G$_0$W-RPA theory.  A complex set of additional solutions to Eq.~(\ref{eqn:poles}) appear
 between the split Dirac point and extend into the lower cone, which correspond to new collective states called plasmarons.  
   (c) $A(k,\omega)$ vs $\omega$ for  $k=k_F$, $k^+$, and 0.  Inset: three dimensional view of $A(k,\omega)$ showing constant momentum cuts (solid color lines) for these three $k$ values. }
\end{figure}

There have been significant advances in infrared nanoscopy which effectively allow for the measurement of the finite $q$ optical conductivity $\sigma(\boldsymbol{q},\omega)$ 
as a function of $\omega$.  A cantilever tip of an atomic force microscope (AFM) operating in a tapping mode (as depicted in Fig.~\ref{fig:1}(a)) allows for the acquisition of 
near field data containing information on the finite $q$ conductivity.   In the paper by Fei et al. \cite{fei:2011} a confinement region of $\approx 30$~nm is achieved 
providing a distribution of $q$ values which peaks around $q=3.4\times10^5$~cm$^{-1}$.  As described by these authors, a super sharp silicon tip
 might reduce the confinement to 10~nm and tips based on carbon nano-tubes to perhaps 1~nm.  As an example, one sees from the ARPES data presented in Fig.~2(b) of 
Bostwick et al.\cite{bostwick:2010} [and shown theoretically here in Fig.~\ref{fig:2}(a)] 
that the Fermi momentum, $k_F\simeq 60\times10^{-3}$\AA$^{-1}$, for $\mu$ of roughly $600$~meV.  
The plasmaron region falls between 500 and 800 meV and corresponds to a momentum region about $k=0$ of roughly $20\times 10^{-3}$\AA$^{-1}$. 
 To access this region a $q$ of order $40\times 10^{-3}$\AA$^{-1}$ is required which is beyond the reach of the tip 
used by Fei et al.\cite{fei:2011} but in the range that could be achieved with a carbon nanotube tip.  However, reducing the doping to obtain a chemical 
potential of $200$~meV would bring the presently available $q$ close to the plasmaron region. Indeed, previous $q=0$ optical experiments\cite{li:2008} were performed on samples 
on SiO$_2$ substrates with
$\mu\sim 150-300$ meV. The scaling of the plasmaron feature with $\mu$ implies that the current range of $q$ values might already be able to access the plasmaron ring
region.

The Kubo formula for the real part of the finite $q$ optical conductivity is given in terms of the universal background conductivity $\sigma_0$ by
\small
\begin{equation}
\frac{\sigma(\boldsymbol{q},\omega)}{\sigma_0}=\frac{8}{\omega}\int\limits_{-\omega}^0 d\omega^\prime \int \frac{d^2\boldsymbol{k}}{2\pi} \sum_{s,s^\prime=\pm} F_{ss^\prime}(\phi) A^s(\boldsymbol{k},\omega^\prime)A^{s^\prime}(\boldsymbol{k}+\boldsymbol{q},\omega^\prime +\omega), \label{eqn:cond}
\end{equation}
\normalsize
where $A^s(\boldsymbol{k},\omega)$ is the spectral function of the Dirac fermions and $F_{ss^\prime}(\phi)=\frac{1}{2}[1+ss^\prime \cos\phi]$.  
Here $\phi=\theta+\beta$, where $\beta= \arctan(\frac{k\sin\theta}{q+k\cos\theta})$ and $\theta$ is the angle between $\boldsymbol{k}$ and $\boldsymbol{q}$.  Note here that $\phi$ differs from that associated with the polarization function, $\Pi(\boldsymbol{q},\omega)$, wherein $\phi=\theta-\beta$.  
While for bare bands with $\boldsymbol{q}\to 0$, $\sigma(\boldsymbol{q},\omega)= i e^2 \omega \Pi(\boldsymbol{q},\omega)/q^2$ as seen in Ref.~\cite{wunsch:2006}, at finite $\boldsymbol{q}$ and with interactions this relation does not hold.
For graphene, the spectral function can be written in terms of the self-energy for the $s=\pm$ cones, $\Sigma_s(\boldsymbol{k},\omega)$, as
\begin{equation}
A^s(\boldsymbol{k},\omega)=\frac{1}{\pi} \frac{|{\rm Im}\Sigma_s(\boldsymbol{k},\omega)|}{[\omega- {\rm Re}\Sigma_s(\boldsymbol{k},\omega)- \epsilon_{{k}}^s]^2 +[{\rm Im}\Sigma_s(\boldsymbol{k},\omega)]^2},\label{eqn:akw}
\end{equation}
where the bare band energies are given by $\epsilon_{{k}}^s=sv_Fk -\mu$.  In this work we use the standard G$_0$W-RPA approximation\cite{leblanc:2011} 
for the self-energy arising from Coulomb interactions (EEI). We have set $\hbar=1$ and will also set the Fermi velocity $v_F=1$ when discussing energy transfer.   
In Fig.~\ref{fig:1}(b) we show the Dirac cones for finite chemical potential $\mu$.  The plasmaron region which we would like to access falls near $k=0$ and extends to 
 $k=k^+$ as shown in Fig.~\ref{fig:2}(a). A crossing occurs between a plasmaron band and a Dirac quasiparticle band in this region.
  The plasmaron structure in Fig.~\ref{fig:2} scales with the value of the chemical potential $\mu$, and Fermi momentum, $k_F$, 
and applies to any $\mu$.  Returning to Fig.~\ref{fig:1}(b)(right), finite ${q}$ optical transitions are shown in which a hole, depicted as an open circle is created by the light 
with the excited final state electron at $k=k_F$, shown as a solid (red) circle.  It is clear from this picture that these new intraband and interband transitions sample states 
in the Dirac cone down to $k=k_F-q$ which approaches the Dirac point at $k=0$ as $q$ approaches the Fermi momentum.  
To be close to, or even inside of, the plasmaron ring region, a momentum of $q\sim k_F-k^+$ is needed.  
The overall effect of finite $q$ probing is to shrink the Pauli-blocked region towards the Dirac point, thus uncovering 
the region of $k$-space where plasmarons are most prominent.  

Plasmaronic modes and dressed Dirac quasiparticles are found in solutions of the equation
\begin{equation}
\omega -{\rm Re}\Sigma_s(\boldsymbol{k},\omega)-\epsilon_{{k}}^s =0. \label{eqn:poles}
\end{equation}
The former are introduced through the interactions and represent new collective modes (or scattering resonance) of a charge carrier plus a plasmon.  
The contours along which Eq.~(\ref{eqn:poles}) is satisfied are indicated by the complicated structure shown in Fig.~\ref{fig:2}(b) with the
upper conical region representing the renormalized Dirac quasiparticle branch and the lower cone, the plasmaronic one.\cite{bostwick:2010}  
This has been probed in detail in ARPES work of Bostwick et al.\cite{bostwick:2010} who also confirm that G$_0$W-RPA theory provides 
a good description of the spectral function of the Dirac fermions in graphene.  A color plot of our own results for the spectral density, $A(\boldsymbol{k},\omega)$, is 
shown in  Fig.~\ref{fig:2}(a) with emphasis on the region near $k=0$.  One of the motivations of this work is to encourage confirmation of the ARPES results using optical methods.  

Another representation of the plasmaron structure is shown in the bottom frame of Fig.~\ref{fig:2} where we provide in the inset a three dimensional plot of the Dirac carrier spectral density. 
Around $k=k_F$ we see a sharp quasiparticle peak with a very diffuse incoherent  sideband extending to large energy but always flat and small in amplitude  
for $\omega >-2\mu$, as is made clearer in  the solid (black) curve of Fig.~\ref{fig:2}(c).  
Below $\omega \sim -2\mu$ there is a plasmaron peak around $-2.8\mu$ with a small sideband from Dirac quasiparticles 
but this appears at large negative energies well within the region of 
the universal background in optics therefore not of primary interest here.
As $k$ is reduced the plasmaron structure becomes more prominent in the energy window of interest and the dash-dotted (blue) curve for $k=k^+$ shows three
peaks: the main quasiparticle peak near $-0.9\mu$, a plasmaron peak at  $\sim -1.8\mu$ and a peak related to the plasmaron ring (the crossing of
the plasmaron and main quasiparticle bands) in between. 
By $k=0$, two peaks of almost equal magnitude remain in the dashed (red) curve, the upper one marks the quasiparticle Dirac point crossing and
the lower, the plasmaron Dirac point.  It is this rich structure from $\sim k^+$ towards $k=0$ which would be exciting to access with optics.  
However, there does not exist a sharp transition in the evolution of $A(k,\omega)$ as a function of $k$ at $k=k^+$, 
but rather the transition is gradual and important structure representative of electron-electron correlations comes into play before $k=k^+$ and 
at energies larger than $\simeq -2.0 \mu$ and this is also of great interest.

\begin{figure}
  \begin{center}
  \includegraphics[width=0.8\linewidth]{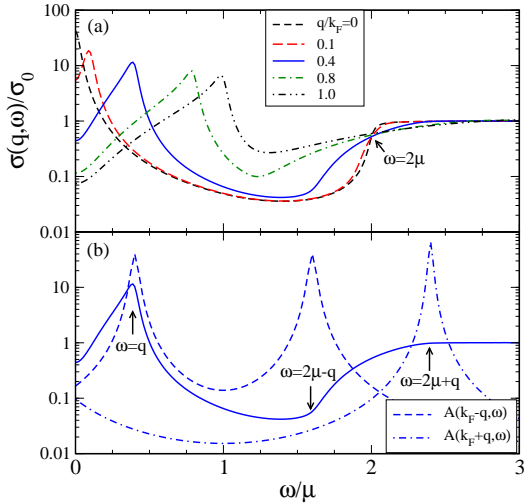}
  \end{center}
  \caption{\label{fig:4}(Color online) (a) The real part of the finite momentum optical conductivity $\sigma(\boldsymbol{q},\omega)$ as a function of $\omega/\mu$ 
for $q/k_F=0$, 0.1, 0.4, 0.8 and 1.0 for bare bands.  Prominent quasiparticle peaks at $\omega=q$ are identified.  The interband edge which rises 
sharply at $\omega=2\mu$ for $q=0$ becomes progressively less prominent with increasing $q$ and shifts to $2\mu-q$.  Another feature is seen at $\omega=2\mu+q$ below which the 
universal background is depleted for finite $q$.   (b) The finite $q$ optical conductivity $\sigma(q,\omega)$ for $q$=0.4,  
compared with the spectral function $A(k,\omega)$ for $k=k_F-q$ and $k=k_F+q$.  }
\end{figure}

\begin{figure}
  \begin{center}
  \includegraphics[width=0.8\linewidth]{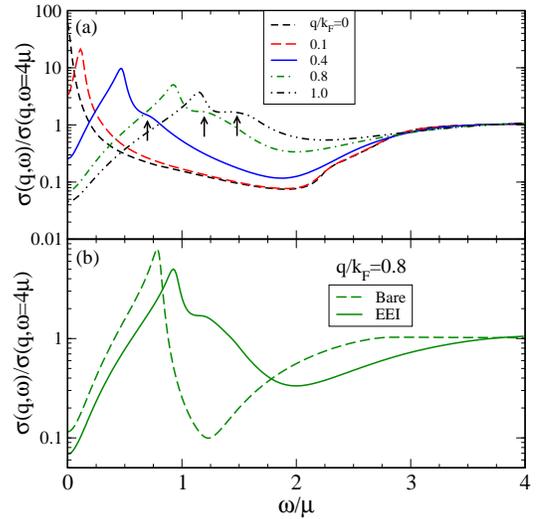}
  \end{center}
  \caption{(Color online) \label{fig:5} (a) The real part of $\sigma(q,\omega)$ for $q/k_F$=0.0, 0.1, 0.4, 0.8, and 1.0 including EEI.  
Quasiparticle peaks are seen in each curve but displaced from their bare band positions
and these are followed by new plasmaronic shoulders indicated by arrows. Interactions 
increase the filling of the Pauli-blocked region as emphasized in frame (b) where interacting and
non-interacting cases are compared.}
\end{figure}

In Fig.~\ref{fig:4}, we present our results for the finite $q$ optical conductivity $\sigma(\boldsymbol{q},\omega)$ vs $\omega$ for five values of $q/k_F=0.0$, 0.1, 0.4, 0.8 and 1.0.
The curves are for the case of no EEI.   In the numerical work, we have used a constant residual scattering rate, $\gamma$ of $0.015$.
  The solid black curve for $q=0$ is well known.  It has a Drude peak centered about $\omega=0$ which decays rapidly according to the value of $\gamma$.  The region of near zero 
conductivity is followed with a sharp rise towards the universal value $\sigma_0$ at $\omega \geq 2\mu$.  As $q$ is increased $\sigma(q,\omega)$ shows a new peak at $\omega=q$ (recall, $v_F=1$ here)
 corresponding to an average of displaced oscillators with $\omega= \epsilon_{\boldsymbol{k}+\boldsymbol{q}}^+ - \epsilon_{\boldsymbol{k}}^+$ which has a maximum value of $q$.  
This is followed by a drop in the conductivity which becomes progressively less as $q$ is increased.  
There is a point of inflection at $\omega=2\mu-q$, the energy at which the interband transitions begin [see Fig.~\ref{fig:1}(c)].  Finite $q$ provides absorption in the region 
which would be Pauli-blocked for $q=0$ and this increases with increasing $q$.  Another important energy scale is $2\mu+q$ above which the universal plateau shows no 
depletion and below which there is some reduction.  All the curves meet at a single point  $\omega=2\mu$ and this crossing could be used to identify the chemical potential.  
The first peaks in each of the curves for $\sigma(q,\omega)$ vs $\omega$ is correlated directly with the quasiparticle peak in the spectral density, 
$A(k-q,\omega)$ (dashed blue curve) of Fig.~\ref{fig:4}(b).  The spectral density is the sum of contributions 
from the top and bottom Dirac cones so that there are two peaks in $A(k,\omega)$.  The second peak at high frequency correlates with the onset of the interband transitions.  
The dashed-dotted curve is $A({k}+{q},\omega)$ and this peak is at $\omega=2\mu+q$ where there is a kink in the conductivity.
These curves are to be compared with those in Fig.~\ref{fig:5}(a) which have the same format as the bare band results but include the EEI in Eq.~(\ref{eqn:akw}).  
Each curve displays a clear quasiparticle peak displaced from the bare case as a result of the appearance of the real part of the 
self-energy in Eq.~(\ref{eqn:poles}).  More importantly, new structure (identified by arrows) appears in $\sigma(q,\omega)$, not seen in the bare band case and 
are due to correlation.  
While such effects are, in principle, also present in the $q=0$ case, they provide only a small modulations\cite{jungseek:2012} on the shape of the intraband Drude peak 
which depends only on electrons with $k=k_F$.  As seen in the inset of Fig.~\ref{fig:2}(c), correlation sidebands are very weak for these electrons.
  But EEI structure is sufficiently sharp to show clearly in the $q/k_F=0.4$ curve.  
This comes from the plasmaron sideband shown in Fig.~\ref{fig:2}(a) which grows in amplitude and sharpens as $k$ is reduced from $k=k_F$ to $k=0$.  
The optical conductivity for $q/k_F=1$ indeed samples the bands converging to the two split Dirac points at $k=0$ and in this case the first peak around $\omega\sim 1.1\mu$
is the Dirac point from the main quasiparticle bands (shifted slightly from $\omega=\mu$ by the renormalizations)
and the second hump  around $\omega\sim 1.6\mu$ is due to the plasmaron band Dirac crossing.
To make our point clear we plot the positions of the peaks in Fig.~\ref{fig:5}(a) for each $q$ on the band structure of Fig.~\ref{fig:2}(a) using $\omega \to -\omega$ and $|k|=|k_F-q|$.  These are shown as open triangles (circles) and track the plasmaron (quasiparticle) band.  This demonstrates how measurements from optics could be used to trace the renormalized band structure.
In Fig.~\ref{fig:5}(b), the dashed green curve is reproduced from Fig.~\ref{fig:4}(a) and is shown for easy comparison with the solid green curve for the same $q/k_F=0.8$
which includes EEI.  
Not only is the quasiparticle peak shifted and a plasmaron shoulder seen, but interactions have filled in the large dip present in the case of the bare bands

In summary, for the bare bands the intraband optical transitions, which provided a Drude peak in $\sigma(q,\omega)$ for $q=0$, are now, for finite $q$ values, distributed over an energy range 
with the dominant peak moved to $\omega=q$.  This fills the Pauli blocked region of the $q=0$ case.  Furthermore, the interband transitions now start at $\omega=2\mu-q$ rather 
than $2\mu$, and the universal background $\sigma_0$ shows some depletion up to $2\mu+q$.  Electron-electron interactions shift the bare band peak at $\omega=q$ to higher energies, 
reflecting quasiparticle energy renormalizations.  They also provide further filling of the Pauli-blocked region and show easily identifiable 
new shoulder-like structures which have their origin in the plasmaronic side bands of the Dirac fermion spectral density, $A({k},\omega)$.  The signature of 
these modes becomes progressively more prominent as the region about $k=0$ is sampled.


\begin{acknowledgments} We thank D.~N. Basov for helpful discussion.
This research was supported in part by NSERC, CIFAR and the NSF under Grant No. NSF PHY05-51164.  
\end{acknowledgments}


\bibliographystyle{apsrev}
\bibliography{bib}

\begin{thebibliography}{25}
\expandafter\ifx\csname natexlab\endcsname\relax\def\natexlab#1{#1}\fi
\expandafter\ifx\csname bibnamefont\endcsname\relax
  \def\bibnamefont#1{#1}\fi
\expandafter\ifx\csname bibfnamefont\endcsname\relax
  \def\bibfnamefont#1{#1}\fi
\expandafter\ifx\csname citenamefont\endcsname\relax
  \def\citenamefont#1{#1}\fi
\expandafter\ifx\csname url\endcsname\relax
  \def\url#1{\texttt{#1}}\fi
\expandafter\ifx\csname urlprefix\endcsname\relax\def\urlprefix{URL }\fi
\providecommand{\bibinfo}[2]{#2}
\providecommand{\eprint}[2][]{\url{#2}}

\bibitem[{\citenamefont{Bostwick et~al.}(2010)\citenamefont{Bostwick, Speck,
  Seyller, Horn, Polini, Asgari, Macdonald, and Rotenberg}}]{bostwick:2010}
\bibinfo{author}{\bibfnamefont{A.}~\bibnamefont{Bostwick}},
  \bibinfo{author}{\bibfnamefont{F.}~\bibnamefont{Speck}},
  \bibinfo{author}{\bibfnamefont{T.}~\bibnamefont{Seyller}},
  \bibinfo{author}{\bibfnamefont{K.}~\bibnamefont{Horn}},
  \bibinfo{author}{\bibfnamefont{M.}~\bibnamefont{Polini}},
  \bibinfo{author}{\bibfnamefont{R.}~\bibnamefont{Asgari}},
  \bibinfo{author}{\bibfnamefont{A.~H.} \bibnamefont{Macdonald}},
  \bibnamefont{and}
  \bibinfo{author}{\bibfnamefont{E.}~\bibnamefont{Rotenberg}},
  \bibinfo{journal}{Science} \textbf{\bibinfo{volume}{328}},
  \bibinfo{pages}{999} (\bibinfo{year}{2010}).

\bibitem[{\citenamefont{Hwang and {Das Sarma}}(2008)}]{hwang:2008}
\bibinfo{author}{\bibfnamefont{E.~H.} \bibnamefont{Hwang}} \bibnamefont{and}
  \bibinfo{author}{\bibfnamefont{S.}~\bibnamefont{{Das Sarma}}},
  \bibinfo{journal}{Phys. Rev. B} \textbf{\bibinfo{volume}{77}},
  \bibinfo{pages}{081412(R)} (\bibinfo{year}{2008}).

\bibitem[{\citenamefont{Polini et~al.}(2008)\citenamefont{Polini, Asgari,
  Borghi, Barlas, Pereg-Barnea, and MacDonald}}]{polini:2008}
\bibinfo{author}{\bibfnamefont{M.}~\bibnamefont{Polini}},
  \bibinfo{author}{\bibfnamefont{R.}~\bibnamefont{Asgari}},
  \bibinfo{author}{\bibfnamefont{G.}~\bibnamefont{Borghi}},
  \bibinfo{author}{\bibfnamefont{Y.}~\bibnamefont{Barlas}},
  \bibinfo{author}{\bibfnamefont{T.}~\bibnamefont{Pereg-Barnea}},
  \bibnamefont{and} \bibinfo{author}{\bibfnamefont{A.~H.}
  \bibnamefont{MacDonald}}, \bibinfo{journal}{Phys. Rev. B}
  \textbf{\bibinfo{volume}{77}}, \bibinfo{pages}{081411(R)}
  (\bibinfo{year}{2008}).

\bibitem[{\citenamefont{Sensarma et~al.}(2011)\citenamefont{Sensarma, Hwang,
  and {Das Sarma}}}]{sensarma:2011}
\bibinfo{author}{\bibfnamefont{R.}~\bibnamefont{Sensarma}},
  \bibinfo{author}{\bibfnamefont{E.~H.} \bibnamefont{Hwang}}, \bibnamefont{and}
  \bibinfo{author}{\bibfnamefont{S.}~\bibnamefont{{Das Sarma}}},
  \bibinfo{journal}{Phys. Rev. B} \textbf{\bibinfo{volume}{84}},
  \bibinfo{pages}{041408(R)} (\bibinfo{year}{2011}).

\bibitem[{\citenamefont{Hwang and {Das Sarma}}(2007)}]{hwang:2007}
\bibinfo{author}{\bibfnamefont{E.~H.} \bibnamefont{Hwang}} \bibnamefont{and}
  \bibinfo{author}{\bibfnamefont{S.}~\bibnamefont{{Das Sarma}}},
  \bibinfo{journal}{Phys. Rev. B} \textbf{\bibinfo{volume}{75}},
  \bibinfo{pages}{205418} (\bibinfo{year}{2007}).

\bibitem[{\citenamefont{Sensarma et~al.}(2010)\citenamefont{Sensarma, Hwang,
  and {Das Sarma}}}]{sensarma:2010}
\bibinfo{author}{\bibfnamefont{R.}~\bibnamefont{Sensarma}},
  \bibinfo{author}{\bibfnamefont{E.~H.} \bibnamefont{Hwang}}, \bibnamefont{and}
  \bibinfo{author}{\bibfnamefont{S.}~\bibnamefont{{Das Sarma}}},
  \bibinfo{journal}{Phys. Rev. B} \textbf{\bibinfo{volume}{82}},
  \bibinfo{pages}{195428} (\bibinfo{year}{2010}).

\bibitem[{\citenamefont{Lundqvist}(1967)}]{lundqvist:1967}
\bibinfo{author}{\bibfnamefont{B.}~\bibnamefont{Lundqvist}},
  \bibinfo{journal}{Phys. Kondens. Materie} \textbf{\bibinfo{volume}{6}},
  \bibinfo{pages}{193} (\bibinfo{year}{1967}).

\bibitem[{\citenamefont{Zhang et~al.}(2008)\citenamefont{Zhang, Brar, Wang,
  Girit, Yayon, Panlasigui, Zettl, and Crommie}}]{zhang:NP:2008}
\bibinfo{author}{\bibfnamefont{Y.}~\bibnamefont{Zhang}},
  \bibinfo{author}{\bibfnamefont{V.~W.} \bibnamefont{Brar}},
  \bibinfo{author}{\bibfnamefont{F.}~\bibnamefont{Wang}},
  \bibinfo{author}{\bibfnamefont{C.}~\bibnamefont{Girit}},
  \bibinfo{author}{\bibfnamefont{Y.}~\bibnamefont{Yayon}},
  \bibinfo{author}{\bibfnamefont{M.}~\bibnamefont{Panlasigui}},
  \bibinfo{author}{\bibfnamefont{A.}~\bibnamefont{Zettl}}, \bibnamefont{and}
  \bibinfo{author}{\bibfnamefont{M.~F.} \bibnamefont{Crommie}},
  \bibinfo{journal}{Nature Physics} \textbf{\bibinfo{volume}{4}},
  \bibinfo{pages}{627} (\bibinfo{year}{2008}).

\bibitem[{\citenamefont{Brar et~al.}(2010)\citenamefont{Brar, Wickenburg,
  Panlasigui, Park, Wehling, Zhang, Decker, \c{C}a\u{g}lar Girit, Balatsky,
  Louie et~al.}}]{brar:2010}
\bibinfo{author}{\bibfnamefont{V.~W.} \bibnamefont{Brar}},
  \bibinfo{author}{\bibfnamefont{S.}~\bibnamefont{Wickenburg}},
  \bibinfo{author}{\bibfnamefont{M.}~\bibnamefont{Panlasigui}},
  \bibinfo{author}{\bibfnamefont{C.-H.} \bibnamefont{Park}},
  \bibinfo{author}{\bibfnamefont{T.~O.} \bibnamefont{Wehling}},
  \bibinfo{author}{\bibfnamefont{Y.}~\bibnamefont{Zhang}},
  \bibinfo{author}{\bibfnamefont{R.}~\bibnamefont{Decker}},
  \bibinfo{author}{\bibnamefont{\c{C}a\u{g}lar Girit}},
  \bibinfo{author}{\bibfnamefont{A.~V.} \bibnamefont{Balatsky}},
  \bibinfo{author}{\bibfnamefont{S.~G.} \bibnamefont{Louie}},
  \bibnamefont{et~al.}, \bibinfo{journal}{Phys. Rev. Lett.}
  \textbf{\bibinfo{volume}{104}}, \bibinfo{pages}{036805}
  (\bibinfo{year}{2010}).

\bibitem[{\citenamefont{Nicol and Carbotte}(2009)}]{nicol:2009}
\bibinfo{author}{\bibfnamefont{E.~J.} \bibnamefont{Nicol}} \bibnamefont{and}
  \bibinfo{author}{\bibfnamefont{J.~P.} \bibnamefont{Carbotte}},
  \bibinfo{journal}{Phys. Rev. B} \textbf{\bibinfo{volume}{80}},
  \bibinfo{pages}{081415(R)} (\bibinfo{year}{2009}).

\bibitem[{\citenamefont{Carbotte et~al.}(2010)\citenamefont{Carbotte, Nicol,
  and Sharapov}}]{carbotte:2010}
\bibinfo{author}{\bibfnamefont{J.~P.} \bibnamefont{Carbotte}},
  \bibinfo{author}{\bibfnamefont{E.~J.} \bibnamefont{Nicol}}, \bibnamefont{and}
  \bibinfo{author}{\bibfnamefont{S.~G.} \bibnamefont{Sharapov}},
  \bibinfo{journal}{Phys. Rev. B} \textbf{\bibinfo{volume}{81}},
  \bibinfo{pages}{045419} (\bibinfo{year}{2010}).

\bibitem[{\citenamefont{LeBlanc et~al.}(2011)\citenamefont{LeBlanc, Carbotte,
  and Nicol}}]{leblanc:2011}
\bibinfo{author}{\bibfnamefont{J.~P.~F.} \bibnamefont{LeBlanc}},
  \bibinfo{author}{\bibfnamefont{J.~P.} \bibnamefont{Carbotte}},
  \bibnamefont{and} \bibinfo{author}{\bibfnamefont{E.~J.} \bibnamefont{Nicol}},
  \bibinfo{journal}{Phys. Rev. B} \textbf{\bibinfo{volume}{84}},
  \bibinfo{pages}{165448} (\bibinfo{year}{2011}).

\bibitem[{\citenamefont{Principi et~al.}(2011)\citenamefont{Principi, Polini,
  Asgari, and MacDonald}}]{principi:2011}
\bibinfo{author}{\bibfnamefont{A.}~\bibnamefont{Principi}},
  \bibinfo{author}{\bibfnamefont{M.}~\bibnamefont{Polini}},
  \bibinfo{author}{\bibfnamefont{R.}~\bibnamefont{Asgari}}, \bibnamefont{and}
  \bibinfo{author}{\bibfnamefont{A.~H.} \bibnamefont{MacDonald}},
  \bibinfo{journal}{arXiv:cond-mat} p. \bibinfo{pages}{1111.3822v1}
  (\bibinfo{year}{2011}).

\bibitem[{\citenamefont{Shay et~al.}(1971)\citenamefont{Shay, {Johnston, Jr.},
  Buehler, and Wernick}}]{shay:1971}
\bibinfo{author}{\bibfnamefont{J.~L.} \bibnamefont{Shay}},
  \bibinfo{author}{\bibfnamefont{W.~D.} \bibnamefont{{Johnston, Jr.}}},
  \bibinfo{author}{\bibfnamefont{E.}~\bibnamefont{Buehler}}, \bibnamefont{and}
  \bibinfo{author}{\bibfnamefont{J.~H.} \bibnamefont{Wernick}},
  \bibinfo{journal}{Phys. Rev. Lett.} \textbf{\bibinfo{volume}{27}},
  \bibinfo{pages}{711} (\bibinfo{year}{1971}).

\bibitem[{\citenamefont{Tediosi et~al.}(2007)\citenamefont{Tediosi, Armitage,
  Giannini, and {van der Marel}}}]{tediosi:2007}
\bibinfo{author}{\bibfnamefont{R.}~\bibnamefont{Tediosi}},
  \bibinfo{author}{\bibfnamefont{N.~P.} \bibnamefont{Armitage}},
  \bibinfo{author}{\bibfnamefont{E.}~\bibnamefont{Giannini}}, \bibnamefont{and}
  \bibinfo{author}{\bibfnamefont{D.}~\bibnamefont{{van der Marel}}},
  \bibinfo{journal}{Phys. Rev. Lett.} \textbf{\bibinfo{volume}{99}},
  \bibinfo{pages}{016406} (\bibinfo{year}{2007}).

\bibitem[{\citenamefont{Armitage et~al.}(2010)\citenamefont{Armitage, Tediosi,
  L\'evy, Giannini, Forro, and {van der Marel}}}]{armitage:2010}
\bibinfo{author}{\bibfnamefont{N.~P.} \bibnamefont{Armitage}},
  \bibinfo{author}{\bibfnamefont{R.}~\bibnamefont{Tediosi}},
  \bibinfo{author}{\bibfnamefont{F.}~\bibnamefont{L\'evy}},
  \bibinfo{author}{\bibfnamefont{E.}~\bibnamefont{Giannini}},
  \bibinfo{author}{\bibfnamefont{L.}~\bibnamefont{Forro}}, \bibnamefont{and}
  \bibinfo{author}{\bibfnamefont{D.}~\bibnamefont{{van der Marel}}},
  \bibinfo{journal}{Phys. Rev. Lett.} \textbf{\bibinfo{volume}{104}},
  \bibinfo{pages}{237401} (\bibinfo{year}{2010}).

\bibitem[{\citenamefont{Orlita and Potemski}(2010)}]{orlita:2010}
\bibinfo{author}{\bibfnamefont{M.}~\bibnamefont{Orlita}} \bibnamefont{and}
  \bibinfo{author}{\bibfnamefont{M.}~\bibnamefont{Potemski}},
  \bibinfo{journal}{Semiconductor Science Technology}
  \textbf{\bibinfo{volume}{25}}, \bibinfo{pages}{063001}
  (\bibinfo{year}{2010}).

\bibitem[{\citenamefont{Gusynin et~al.}(2007)\citenamefont{Gusynin, Sharapov,
  and Carbotte}}]{gusynin:2007}
\bibinfo{author}{\bibfnamefont{V.~P.} \bibnamefont{Gusynin}},
  \bibinfo{author}{\bibfnamefont{S.~G.} \bibnamefont{Sharapov}},
  \bibnamefont{and} \bibinfo{author}{\bibfnamefont{J.~P.}
  \bibnamefont{Carbotte}}, \bibinfo{journal}{Phys. Rev. Lett.}
  \textbf{\bibinfo{volume}{98}}, \bibinfo{pages}{157402}
  (\bibinfo{year}{2007}).

\bibitem[{\citenamefont{Gusynin et~al.}(2009)\citenamefont{Gusynin, Sharapov,
  and Carbotte}}]{gusynin:2009}
\bibinfo{author}{\bibfnamefont{V.~P.} \bibnamefont{Gusynin}},
  \bibinfo{author}{\bibfnamefont{S.~G.} \bibnamefont{Sharapov}},
  \bibnamefont{and} \bibinfo{author}{\bibfnamefont{J.~P.}
  \bibnamefont{Carbotte}}, \bibinfo{journal}{New Journal of Physics}
  \textbf{\bibinfo{volume}{11}}, \bibinfo{pages}{095013}
  (\bibinfo{year}{2009}).

\bibitem[{\citenamefont{Peres et~al.}(2008)\citenamefont{Peres, Stauber, and
  {Castro Neto}}}]{peres:2008}
\bibinfo{author}{\bibfnamefont{N.~M.~R.} \bibnamefont{Peres}},
  \bibinfo{author}{\bibfnamefont{T.}~\bibnamefont{Stauber}}, \bibnamefont{and}
  \bibinfo{author}{\bibfnamefont{A.~H.} \bibnamefont{{Castro Neto}}},
  \bibinfo{journal}{EPL} \textbf{\bibinfo{volume}{84}}, \bibinfo{pages}{38002}
  (\bibinfo{year}{2008}).

\bibitem[{\citenamefont{Li et~al.}(2008)\citenamefont{Li, Henriksen, Jiang,
  Hao, Martin, Kim, Stormer, and Basov}}]{li:2008}
\bibinfo{author}{\bibfnamefont{Z.}~\bibnamefont{Li}},
  \bibinfo{author}{\bibfnamefont{E.~A.} \bibnamefont{Henriksen}},
  \bibinfo{author}{\bibfnamefont{Z.}~\bibnamefont{Jiang}},
  \bibinfo{author}{\bibfnamefont{Z.}~\bibnamefont{Hao}},
  \bibinfo{author}{\bibfnamefont{M.~C.} \bibnamefont{Martin}},
  \bibinfo{author}{\bibfnamefont{P.}~\bibnamefont{Kim}},
  \bibinfo{author}{\bibfnamefont{H.~L.} \bibnamefont{Stormer}},
  \bibnamefont{and} \bibinfo{author}{\bibfnamefont{D.~N.} \bibnamefont{Basov}},
  \bibinfo{journal}{Nat. Phys.} \textbf{\bibinfo{volume}{4}},
  \bibinfo{pages}{532} (\bibinfo{year}{2008}).

\bibitem[{\citenamefont{Stauber and Peres}(2008)}]{stauber:2008b}
\bibinfo{author}{\bibfnamefont{T.}~\bibnamefont{Stauber}} \bibnamefont{and}
  \bibinfo{author}{\bibfnamefont{N.~M.~R.} \bibnamefont{Peres}},
  \bibinfo{journal}{J. Phys.: Condens. Matter} \textbf{\bibinfo{volume}{20}},
  \bibinfo{pages}{055002} (\bibinfo{year}{2008}).

\bibitem[{\citenamefont{Fei et~al.}(2011)\citenamefont{Fei, Andreev, Bao,
  Zhang, McLeod, Wang, Stewart, Zhao, Dominguez, Thiemens et~al.}}]{fei:2011}
\bibinfo{author}{\bibfnamefont{Z.}~\bibnamefont{Fei}},
  \bibinfo{author}{\bibfnamefont{G.~O.} \bibnamefont{Andreev}},
  \bibinfo{author}{\bibfnamefont{W.}~\bibnamefont{Bao}},
  \bibinfo{author}{\bibfnamefont{L.~M.} \bibnamefont{Zhang}},
  \bibinfo{author}{\bibfnamefont{A.~S.} \bibnamefont{McLeod}},
  \bibinfo{author}{\bibfnamefont{C.}~\bibnamefont{Wang}},
  \bibinfo{author}{\bibfnamefont{M.~K.} \bibnamefont{Stewart}},
  \bibinfo{author}{\bibfnamefont{Z.}~\bibnamefont{Zhao}},
  \bibinfo{author}{\bibfnamefont{G.}~\bibnamefont{Dominguez}},
  \bibinfo{author}{\bibfnamefont{M.}~\bibnamefont{Thiemens}},
  \bibnamefont{et~al.}, \bibinfo{journal}{Nano Letters}
  \textbf{\bibinfo{volume}{11}}, \bibinfo{pages}{4701} (\bibinfo{year}{2011}).

\bibitem[{\citenamefont{Wunsch et~al.}(2006)\citenamefont{Wunsch, Stauber,
  Sols, and Guinea}}]{wunsch:2006}
\bibinfo{author}{\bibfnamefont{B.}~\bibnamefont{Wunsch}},
  \bibinfo{author}{\bibfnamefont{T.}~\bibnamefont{Stauber}},
  \bibinfo{author}{\bibfnamefont{F.}~\bibnamefont{Sols}}, \bibnamefont{and}
  \bibinfo{author}{\bibfnamefont{F.}~\bibnamefont{Guinea}},
  \bibinfo{journal}{New Journal of Physics} \textbf{\bibinfo{volume}{8}},
  \bibinfo{pages}{318} (\bibinfo{year}{2006}).

\bibitem[{\citenamefont{Hwang et~al.}(2012)\citenamefont{Hwang, LeBlanc, and
  Carbotte}}]{jungseek:2012}
\bibinfo{author}{\bibfnamefont{J.}~\bibnamefont{Hwang}},
  \bibinfo{author}{\bibfnamefont{J.~P.~F.} \bibnamefont{LeBlanc}},
  \bibnamefont{and} \bibinfo{author}{\bibfnamefont{J.~P.}
  \bibnamefont{Carbotte}}, \bibinfo{journal}{arXiv:cond-mat} p.
  \bibinfo{pages}{1202.1059v1} (\bibinfo{year}{2012}).

\end{thebibliography}

\end{document}